\documentclass[10pt,conference]{IEEEtran}
\IEEEoverridecommandlockouts

\usepackage{amssymb}
\usepackage[cmex10]{amsmath}
\usepackage{stfloats}
\usepackage{graphicx}
\usepackage{subfigure}
\usepackage{tabularx}
\usepackage{epsfig,epsf,color,balance,cite}
\usepackage{verbatim}
\usepackage{url}
\usepackage{bm}
\usepackage{booktabs}
\usepackage[ruled,linesnumbered]{algorithm2e}
\usepackage{algorithmic}

\usepackage{geometry}
\usepackage{color}
\definecolor{myc1}{rgb}{0,0,0}

\begin{document}

\title{ 
On Performance of Fluid Antenna Relay (FAR)-Assisted AAV-NOMA Wireless Network
\vspace{-0.5cm}}

\author{
\IEEEauthorblockN{
Ruopeng Xu$\IEEEauthorrefmark{1}$,
Songling Zhang$\IEEEauthorrefmark{1}$,
Zhaohui Yang$\IEEEauthorrefmark{1}$,
Yixuan Chen$\IEEEauthorrefmark{1}$,\\
Mingzhe Chen$\IEEEauthorrefmark{2}$,
Zhaoyang Zhang$\IEEEauthorrefmark{1}$,
Kai-Kit Wong$\IEEEauthorrefmark{3}$
}
	\IEEEauthorblockA{
			$\IEEEauthorrefmark{1}$College of Information Science and Electronic Engineering, Zhejiang University, Hangzhou, China\\
            $\IEEEauthorrefmark{2}$Department of Electrical and Computer Engineering, University of Miami \\
            $\IEEEauthorrefmark{3}$Department of Electronic and Electrical Engineering, University College London, U.K.\\
          	E-mails:
(ruopengxu, sl-zhang, yang\_zhaohui, chen\_yixuan, ning\_ming)@zju.edu.cn,\\
mingzhe.chen@miami.edu, kai-kit.wong@ucl.ac.uk
		}
\thanks{This work was supported by the National Natural Science Foundation of China (NSFC) under Grants 62394292, 62394290, Zhejiang Key R\&D Program under Grant 2023C01021, Young Elite Scientists Sponsorship Program by China Association for Science and Technology under Grant 2023QNRC001, the Fundamental Research Funds for the Central Universities under Grant 226-2024-00069.}
\vspace{-3em}
}
\maketitle

\maketitle

\begin{abstract}
In this paper, we investigate the performance of a fluid antenna relay (FAR)-assisted downlink communication system utilizing non-orthogonal multiple access (NOMA). The FAR, which integrates a fluid antenna system (FAS), is equipped on an autonomous aerial vehicle (AAV), and introduces extra degrees of freedom to improve the performance of the system. The transmission is divided into a first phase from the base station (BS) to the users and the FAR, and a second phase where the FAR forwards the signal using amplify-and-forward (AF) or decode-and-forward (DF) relaying to reduce the outage probability (OP) for the user maintaining weaker channel conditions. To analyze the OP performance of the weak user, Copula theory and the Gaussian copula function are employed to model the statistical distribution of the FAS channels. Analytical expressions for weak user's OP are derived for both the AF and the DF schemes. Simulation results validate the effectiveness of the proposed scheme, showing that it consistently outperforms benchmark schemes without the FAR. In addition, numerical simulations also demonstrate the values of the relaying scheme selection parameter under different FAR positions and communication outage thresholds.
\end{abstract}

\IEEEpeerreviewmaketitle

\section{Introduction}\label{Introduction}
The development of sixth-generation (6G) mobile communication systems imposes new requirements on communication network performance\cite{dong2025communication,yang2023energy,2024arXiv241202538Y}. Fluid antenna system (FAS), an emerging technology bringing additional spatial diversity gains to the communication system\cite{wong2020fluid,11319344,xu2024capacity}, has recently been viewed as a promising candidate to help communication systems meet the 6G communication metrics. In particular, the use of diversity techniques can mitigate signal fading in wireless networks\cite{laneman2004cooperative}, thus introducing FAS into existing relay communication systems deployed traditional antenna systems (TASs) has great potential to deal with severe signal attenuation in high-frequency 6G networks and expand the coverage of the communication system. Meanwhile, another approach to enhance the quality of communications in relay networks is the deployment of autonomous aerial vehicles (AAVs) in low-altitude airspace\cite{lyu2023low,wang2023autonomous,jiang2024study} by establishing a high probability of line-of-sight (LoS) links. 

Research works that deploy FAS as a relay\cite{aka2025power,xu2025energy,10167904} or combine FAS with AAVs\cite{abdou2024sum,yang2025rate,xu2025transformer} have previously been investigated. However, existing works primarily consider the single relaying scheme, i.e., using either amplify-and-forward (AF)\cite{aka2025power,xu2025energy} or decode-and-forward (DF)\cite{10167904,abdou2024sum} for relaying, neglecting the need for different relaying schemes under different communication conditions. In particular, DF does not introduce extra noise during forwarding, but is susceptible to error propagation due to decoding process, while AF has no risk of decoding errors, but it amplifies noise, making the usage of a single fixed mode cannot always fit all communication conditions. In addition, some of the existing FAR-AAV relay work employs orthogonal multiple access (OMA) techniques\cite{yang2025rate}. OMA often allocates resources uniformly, leading to poor fairness where weak-channel users struggle to meet the required quality of service (QoS). Moreover, coordination between relay links and direct links from the base station (BS) has not been fully utilized in \cite{yang2025rate,abdou2024sum,xu2025transformer}. 

Motivated by these research gaps, we propose to deploy FAS on a AAV to constitute an FAR relay node and utilize non-orthogonal multiple access (NOMA) coordinately serving users in high outage probability (OP) with only direct links, where the FAR adaptively chooses relaying scheme between AF and DF to obtain a lower OP of the weak-channel user.

The key contributions of this paper include:



\begin{itemize}
    \item We investigate an FAR-assisted two-user downlink communication system, where the FAR integrates the FAS and AAV. The overall downlink transmission consists of the first phase, where a superposed signal is transmitted from BS to users and the FAR, and the second phase from the FAR to ground users. After receiving signals from BS in the first phase, the FAR forwards the signal with either AF or DF for relaying following the principle to obtain a lower OP of the user with a weaker channel (user $2$).
    \item To compare the performance of the proposed system with different relaying schemes, we derive the transmission OP for the user $2$ with the FAR using AF and DF, respectively. Furthermore, we introduce Copular theory to represent statistical distribution of the FAS channels, and utilize the Gaussian copula function to approximate the jointly cumulative distribution function (CDF) of the random variables following above distributions of FAS. 
    \item We evaluate and compare the OP performance of user $2$ under different conditions via numerical simulations. Simulation results validate its effectiveness and show the proposed scheme always outperforms the benchmark schemes. Moreover, we also develop simulations about the results on relaying scheme selection under different positions of FAR and different outage thresholds.
\end{itemize}

\section{System Model}
\begin{figure}[t]
\centering
\includegraphics[width=.8\linewidth]{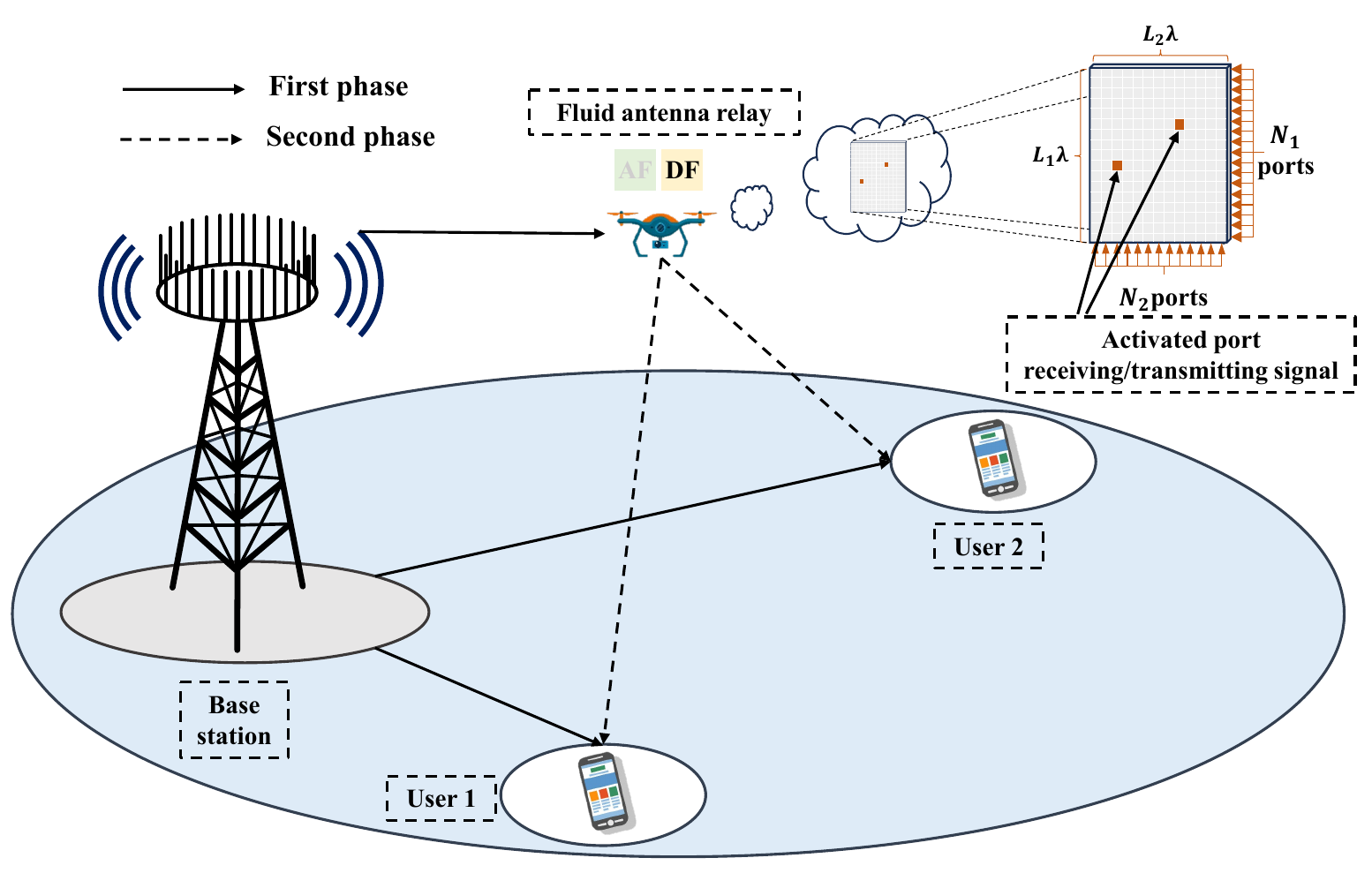}
\caption{System model of the proposed communication.} 
\label{SystemModel}
\vspace{-0.6cm}
\end{figure}

As illustrated in Fig.~\ref{SystemModel}, we consider an FAR-assisted downlink communication with NOMA, which consists of a BS, two ground users, and one AAV. BS and ground users are equipped with a single TAS, and the AAV is equipped with a 2-dimensional (2D) FAS. In particular, the AAV equipped with the FAS is considered as FAR working in half-duplex mode. We assume that the channels between both users and the BS are poor so the communication directly from the BS to each user is in high probability of being outage. Hence, our objective is to deploy the FAR to help establish relay links to improve communications between the BS and the ground users. Moreover, we assume that near user is closer to the BS (user $1$) than the other (user $2$), thus user $1$ has a relatively better channel condition. In the considered model, there are two phases in one downlink transmission. In the first phase, the BS transmits signals to the FAR and ground users, and in the second phase, the FAR transmits the received signals with either AF or DF relaying scheme. 

With activating different ports of FAS, the FAR can obtain different channel gains, and the port acquiring the largest channel gain is the optimal port\cite{wong2020fluid}. Thus, by finding the optimal port, the FAR can greatly improve the communication qualities, while the process also introduces additional overhead for port selection. Without loss of generality, we assume that the FAR receives the signal from the BS with fixed port to save port selection overhead and transmits the signals to ground users by activating the optimal port of FAS to obtain better channel gains and improve communication QoS.  

\subsection{Fluid Antenna Relay Model}
It is assumed that the 2D FAS includes one RF chain and $N=N_1\times N_2$ preset ports, where the $N_i$ ports are uniformly distributed along a linear space of length $L_i\lambda$ for $i\in\{1,2\}$, taking up a grid surface size of $L=L_1\times L_2\lambda^2$ with $\lambda$ being the wave length. By introducing mapping function $\mathcal{F}(\cdot)$, we can denote the $(n_1,n_2)$-th port as the $l$-th port by $l=\mathcal{F}(n_1,n_2)=(n_1-1)N_2+ n_2$.
Based on this notation, we denote the channel gain from FAR to user $j$ with activating the $l$-th port by
\begin{equation}
   h_{{\mathrm{FU}_j}}^l=\sqrt{d_{\mathrm{FU}_j}^{-\alpha}}g_{\mathrm{FU}_j}^l,
\end{equation}
where $g_{\mathrm{FU}_j}^l \sim \mathcal{CN}(0,1) $ is the normalized channel gain using port $l$, following a complex Gaussian distribution with zero mean and unit variance. Then, the square of the amplitude of $h_{\mathrm{FU}_j}^l$ follows the exponential distribution with the parameter $d_{\mathrm{FU}_j}^{\alpha}$, i.e., $|h_{\mathrm{FU}_j}^l|^2\sim \mathrm{EXP}(d_{\mathrm{FU}_j}^\alpha)$. Moreover, the amplitude of the optimal port $h_{\mathrm{FU}_j}$ can be mathematically given as
\begin{equation}\label{h_BF}
    |h_{\mathrm{FU}_j}|^2  = \mathop{\mathrm{max}} \{|h_{{\mathrm{FU}_j}}^1|^2,\dots,|h_{\mathrm{FU}_j}^N|^2\}.
\end{equation}

\subsection{Communication Model}
In downlink transmission, the BS transmits a superposed signal $x$ with transmit power $P_\mathrm{B}$ to the FAR and two ground users in the first transmission phase. The superposed signal consists of the desired symbols of the two users, where $x_i$ is intended to the user $i$, $i=\{1,2\}$. By performing NOMA, the transmit power $P_\mathrm{B}$ is divided by the ratio $\beta_\mathrm{B}$ between the two symbols. Thus, the transmitted signal $x$ can be expressed as
\begin{equation}
    x = \sqrt{\beta_\mathrm{B}P_\mathrm{B}}x_1+\sqrt{(1-\beta_\mathrm{B})P_\mathrm{B}}x_2,
\end{equation}
where $x_i\sim \mathcal{CN}(0,1)$. Then, signal received at the FAR, user $1$ and user $2$ in the first phase can be respectively given as
\begin{equation}
    y_i^{1} = \sqrt{d_{\mathrm{B}i}^{-\alpha}} g_{\mathrm{\mathrm{B}}i}x+n_i^1=h_{\mathrm{B}i}x + n_i^1,\ i=\{\mathrm{F,U_1,U_2}\},
\end{equation}
where the superscript $1$ represents the first transmission phase, and $n_i$ is the additional white Gaussian noise (AWGN). Without loss of generality, we assume AWGN at all terminals maintains the variance $\sigma^2$. 

According to the principles of NOMA, since user $1$ holds a stronger channel, power allocation ratio $\beta_{\mathrm{B}}$ should satisfies that $0 \leq\beta_{\mathrm{B}}<\frac{1}{2} $, and user $2$ decodes its own message with treating the message of user $1$ as the noise. As a result, with defining $\rho_\mathrm{B}=P_\mathrm{B}/\sigma^2$, the signal-to-interference-plus-noise ratio (SINR) for user $2$ to decode its own message is given as
\begin{equation}\label{SINR U2 1}
    \gamma_{\mathrm{U_2}}^1 = \frac{d_{\mathrm{BU_2}}^{-\alpha}|g_{\mathrm{BU}_2}|^2(1-\beta_\mathrm{B})\rho_\mathrm{B}}{d_{\mathrm{BU_2}}^{-\alpha}|g_{\mathrm{BU}_2}|^2\beta_{\mathrm{B}}\rho_\mathrm{B}+1}.
\end{equation}

\subsection{AF Relaying}
If use AF for relaying, FAR receives the signal from the BS and transmits it with an amplify coefficient $\eta$ to two ground users, respectively. In particular, we set the transmit power at the FAR as $P_{\mathrm{F}}$ and the amplify coefficient $\eta$ can be set as $ \eta =  \sqrt{\frac{\rho_\mathrm{F}}{d_{\mathrm{BF}}^{-\alpha}|g_{\mathrm{BF}}|^2\rho_{\mathrm{B}}+1}}$, where $\rho_{\mathrm{F}}=\frac{P_{\mathrm{F}}}{\sigma^2}$.
Then, the received signal in the second transmission phase at user $i$ can be given as
\begin{equation}
    y_{\mathrm{U}_i}^{11} 
    =\sqrt{d_{\mathrm{FU}_i}^{-\alpha}} g_{\mathrm{FU}_i}\eta \left(x\sqrt{d_{\mathrm{BF}}^{-\alpha}}g_{\mathrm{BF}}+n_i^1\right)+n_i^{11}
    , 
\end{equation}
where the superscript $11$ stands for the second transmission phase, and $n_i^{11}$ is the corresponding AWGN at user $i$.

In the second transmission phase, both users should first decode the message of user $2$. Hence, the SINR of user $2$ decoding the message of user $2$ can be given as
\begin{align}\label{SINR U1U2 21}
    \gamma_{\mathrm{U}_2}^{11} = \frac{|h_{\mathrm{BF}}|^2|h_{\mathrm{FU}_2}|^2(1-\beta_\mathrm{B})\rho_{\mathrm{B}}\rho_{\mathrm{F}}}{|h_{\mathrm{BF}}|^2|h_{\mathrm{FU}_2}|^2\beta_\mathrm{B}\rho_{\mathrm{B}}\rho_{\mathrm{F}}+|h_{\mathrm{BF}}|^2\rho_{\mathrm{B}}+|h_{\mathrm{FU}_2}|^2\rho_{\mathrm{F}}+1}.
\end{align}
When $\gamma_{\mathrm{U}_2}^{11}\geq \gamma_{\mathrm{U}_2}$, user $2$ successfully decodes its message.

Based on \eqref{SINR U2 1} and \eqref{SINR U1U2 21}, user $2$ combines the received signals $y_{\mathrm{U}_2}^1$ and $y_{\mathrm{U}_2}^{11}$ to decode its message with the SINR as $ \Gamma_{\mathrm{U}_2}^{\mathrm{AF}} = \gamma_{\mathrm{U_2}}^1 + \gamma_{\mathrm{U_2}}^{11}$, where the superscript $\mathrm{AF}$ is used to show it is the combined SINR using AF relaying scheme.

\subsection{DF Relaying}
If use DF for relaying, the FAR first decodes the received signal from the BS. The SINR and SNR for the FAR to decode $x_2$ and $x_1$ can be given as
\begin{align}\label{FAR U2}
    \gamma_{\mathrm{FU_2}}^1 = \frac{d_{\mathrm{BF}}^{-\alpha}|g_{\mathrm{BF}}|^2(1-\beta_\mathrm{B})\rho_\mathrm{B}}{d_{\mathrm{BF}}^{-\alpha}|g_{\mathrm{BF}}|^2\beta_{\mathrm{B}}\rho_\mathrm{B}+1},
\end{align}
and
\begin{equation}\label{FAR U1}
     \gamma_{\mathrm{FU}_1}^1 = d_{\mathrm{BF}}^{-\alpha}|g_{\mathrm{BF}}|^2\beta_{\mathrm{B}}\rho_{\mathrm{B}},
\end{equation}
respectively. Based on the decoding results, if the FAR can decode $x_1$ successfully, the FAR is able to transmit the superimposed signal with a new power allocation scheme, where the transmitted signal can be expressed as
\begin{equation}
    x_{\mathrm{F}} = \sqrt{\beta_\mathrm{F}P_\mathrm{F}}x_1+\sqrt{(1-\beta_\mathrm{F})P_\mathrm{F}}x_2,
\end{equation}
where $P_\mathrm{F}$ is the transmit power at the FAR and $\beta_{\mathrm{F}}$ is the power allocation coefficient at the FAR with $\beta_{\mathrm{F}}\in[0,\frac{1}{2})$ following NOMA principles. 

Then, for user $2$, the received signal can be denoted by $y_{\mathrm{U}_2}^{11-\mathrm{F}}$, where the superscript $-\mathrm{F}$ emphasizes that the received signal was decoded by the FAR. Moreover, user $2$ decoding $x_2$ after receiving the signal from the FAR obtains the SINR
\begin{equation}\label{SINR U2 U2 F}
    \gamma_{\mathrm{U}_2}^{11-\mathrm{F}} = \frac{d_{\mathrm{FU}_2}^{-\alpha}|g_{\mathrm{FU}_2}|^2(1-\beta_\mathrm{F})\rho_\mathrm{F}}{d_{\mathrm{FU}_2}^{-\alpha}|g_{\mathrm{FU}_2}|^2\beta_{\mathrm{F}}\rho_\mathrm{F}+1}.
\end{equation}

In this case, user $2$ combines $y_{\mathrm{U}_2}^1$ and $y_{\mathrm{U}_2}^{11-\mathrm{F}}$ to decode $x_2$ with the SINR as $\Gamma_{\mathrm{U}_2}^{\mathrm{DF-F}}= \gamma_{\mathrm{U}_2\mathrm{U}_2}^1+\gamma_{\mathrm{U}_2\mathrm{U}_2}^{11-\mathrm{F}}$. 

On the other hand, when the FAR fails to decode $x_1$, it transmits the decoded $x_2$ to the ground users, where the transmitted signal can be given as $x_{\mathrm{F}} = \sqrt{P_{\mathrm{F}}}x_2$. For user $i$, the received signal can be denoted as $y_{\mathrm{U}_i}^{11-\mathrm{2}}$, where the superscript $-\mathrm{2}$ emphasizes only $x_2$ was transmitted by the FAR. The SNR of user $i$ to decode the received signal can be expressed as 
\begin{equation}\label{SINR U2 U2 2}
    \gamma_{\mathrm{U}_2}^{11-2} = d_{\mathrm{FU}_2}^{-\alpha}|g_{\mathrm{FU}_2}|^2\rho_{\mathrm{F}}, 
\end{equation}

In this situation, user $2$ combines $y_{\mathrm{U}_2}^1$ and $y_{\mathrm{U}_2}^{11-2}$ with SINR $\Gamma_{\mathrm{U}_2}^{\mathrm{DF-2}}=\gamma_{\mathrm{U}_2\mathrm{U}_2}^1+\gamma_{\mathrm{U}_2\mathrm{U}_2}^{11-2}$
to decode its own message.

\section{Outage Probability Performance Analysis}\label{OPAnalysis}

\subsection{CDF of FAS with Copula Theory}
For a 2D FAS, the entry of spatial correlation matrix $\mathbf{J}$ of the FAS representing the spatial correlation between the $(n_1,n_2)$-th port and $(\tilde{n}_1,\tilde{n}_2)$-th port can be described as\cite{10303274}
\begin{align}
\nonumber
    &J_{(n_1,n_2),(\tilde{n}_1,\tilde{n}_2)}\\
    &=j_0\left(2\pi\sqrt{\left(\frac{|n_1-\tilde{n}_1|}{N_1-1}L_1\right)^2+\left(\frac{|n_2-\tilde{n}_2|}{N_2-1}L_2\right)^2}\right),
\end{align}
where $j_0(\cdot)$ is the spherical Bessel function of the first kind.  

For the link between the FAR and user $2$, we introduce Copula theory\cite{10.5555/1952073} to acquire the CDF of $|h_{\mathrm{FU}_2}|^2$, and specifically exploit the Gaussian copula function\cite{10678877} to approximate its numerical value. By exploiting the analytical results in \cite{10678877}, $F_{|h_{\mathrm{FU}_2}|^2}\left(x\right)$ can be presented as
\begin{align}\label{GaussionCopula}
\nonumber
&F_{|h_{\mathrm{FU}_2}|^2}\left(x\right) \triangleq \Phi_{\mathbf{J},2}(x) \\
&=\Phi_{\mathbf{J}}\left(\phi^{-1}\left(F_{\left|h_{\mathrm{FU}_2}^{1}\right|^2}\left(x\right)\right),\dots,\phi^{-1}\left(F_{\left|h_{\mathrm{FU}_2}^N\right|^2}\left(x\right)\right)\right),
\end{align}
where $F_{|h_{\mathrm{FU}_2}^i|^2}(x) = 1-e^{-d_{\mathrm{FU}_2}^\alpha x}$, $\Phi_\mathbf{J}(\cdot)$ is the joint CDF of the multivariate normal distribution with zero mean vector and correlation matrix $\mathbf{J}$, $\phi^{-1}(\cdot)$ is the quantile function of the standard normal distribution, i.e., $\phi^{-1}(x)=\sqrt{2}\ \mathrm{erf}^{-1}(2x-1)$, in which $\mathrm{erf}^{-1}$ is the inverse function of error function $\mathrm{erf}(x) = \frac{2}{\sqrt{\pi}}\int_0^{x}e^{-r^2} \mathrm{d}r$.

\subsection{OP with AF Relaying}
For the sake of notation simplicity, $|h_{\mathrm{\mathrm{BU}_1}}|^2$, $|h_{\mathrm{\mathrm{BU}_2}}|^2$, $|h_{\mathrm{\mathrm{BF}}}|^2$ and $|h_{\mathrm{\mathrm{FU}_2}}|^2$ are denoted as independent RVs $W_1$, $W_2$, $X$ and $Z$, respectively.
If use AF for relaying and define $\xi_{\mathrm{B}} = 1-\beta_{\mathrm{B}}-\beta_{\mathrm{B}}\gamma_{\mathrm{U}_2}$, the OP of the user $2$ can be expressed as
\begin{align}
\nonumber
    &q^{\mathrm{AF}} = \mathbb{P}(\Gamma_{\mathrm{U}_2}^{\mathrm{AF}}<\gamma_{\mathrm{U}_2}) =\mathbb{P} \left( \frac{W_2(1-\beta_{\mathrm{B}})\rho_{\mathrm{B}}}{W_2 \beta_{\mathrm{B}}\rho_{\mathrm{B}}} \right. \\
    \nonumber
    &\left. + \frac{XZ(1-\beta_{\mathrm{B}})\rho_{\mathrm{B}}\rho_{\mathrm{F}}}{XZ\beta_{\mathrm{B}}\rho_{\mathrm{B}}\rho_{\mathrm{F}}+X\rho_{\mathrm{B}}+Z\rho_{\mathrm{F}}+1}<\gamma_{\mathrm{U}_2}\right)\\
      \nonumber
    & = 
    \mathbb{P} \bigg( Z\rho_{\mathrm{F}}\Big\{X\rho_{\mathrm{B}}\big[W_2\rho_{\mathrm{B}}\beta_{\mathrm{B}}(\xi_{\mathrm{B}}+1-\beta_{\mathrm{B}})+\xi_{\mathrm{B}}\big]\\
    &+W_2\rho_{\mathrm{B}}\xi_{\mathrm{B}} -\gamma_{\mathrm{U}_2}\Big\} <(\gamma_{\mathrm{U}_2}-W_2\rho_{\mathrm{B}}\xi_{\mathrm{B}})(1+X\rho_{\mathrm{B}})\bigg).
\end{align}

Define $C_{w_2}=\frac{\gamma_{\mathrm{U}_2}}{\xi_{\mathrm{B}}\rho_{\mathrm{B}}}$, ${C_{x}=\frac{\gamma_{\mathrm{U}_2}-\xi_{\mathrm{B}} w_2 \rho_{\mathrm{B}}}{\rho_{\mathrm{B}}[\xi_{\mathrm{B}}+ w_2 \beta_{\mathrm{B}}\rho_{\mathrm{B}}(1+\xi_{\mathrm{B}}-\beta_{\mathrm{B}})]}}$, $C_z=\frac{(\gamma_{\mathrm{U}_2}-w_2\rho_{\mathrm{B}}\xi_{\mathrm{B}})(1+x\rho_{\mathrm{B}})}{\rho_{\mathrm{F}}\{x\rho_{\mathrm{B}}[w_2\rho_{\mathrm{B}\beta_{\mathrm{B}}}(\xi_{\mathrm{B}}+1-\beta_{\mathrm{B}})+\xi_{\mathrm{B}}]+w_2\rho_{\mathrm{B}}\xi_{\mathrm{B}}-\gamma_{\mathrm{U}_2}\}}$, and $C_0 = \frac{-\xi_{\mathrm{B}}}{\rho_{\mathrm{B}}\beta_{\mathrm{B}}(\xi_{\mathrm{B}}+1-\beta_{\mathrm{B}})}$. When $\xi_{\mathrm{B}}>0$, i.e., $\beta_{\mathrm{B}}<\frac{1}{1+\gamma_{\mathrm{U}_2}}$, the transmission will always not outage with $W_2 \geq C_{w_2}$, since the term on the right side is non-positive while the left-hand-side term is non-negative. With $W_2$ being less than $C_{w_2}$, the term on the left of the equality is positive only when $X \geq C_x$ and $Z \geq C_z$. Therefore, when $\beta_{\mathrm{B}}>\frac{1}{1+\gamma_{\mathrm{U}_2}}$, the OP is
    \begin{align}\label{q1 AF}
    \nonumber
        q_1^{\mathrm{AF}} 
       = 1- &e^{-d_{\mathrm{BU}_2}^{\alpha} C_{w_2}} -\int\limits_{0}^{C_{w_2}} \int\limits_{C_x}^{\infty}\big(1- \Phi_{\mathbf{J},2}(C_z) \big)\\
        &\times \ d_{\mathrm{BF}}^{\alpha}e^{-d_{\mathrm{BF}}^\alpha{\mathit{x}}}\ dxd_{\mathrm{BU}_2}^{\alpha}e^{-d_{\mathrm{BU}_2}^\alpha{\mathit{w}_2}}\ d{w_2}.
    \end{align}

When $\xi_{\mathrm{B}}\leq0$, we can find that if $\xi_{\mathrm{B}}+1-\beta_{\mathrm{B}} \leq 0$, i.e., $\beta_{\mathrm{B}}\geq \frac{2}{2+\gamma_{\mathrm{U}_2}}$, all terms on the left hand of the inequality are non-positive, while the term on the right hand side is positive. Hence, the transmission will always outage. On the other hand, when $\frac{1}{1+\gamma_{\mathrm{U}_2}}\leq \beta_{\mathrm{B}} < \frac{2}{2+\gamma_{\mathrm{U}_2}}$, we need to promise $W_2 \geq C_0$ and $X\geq C_x$ to ensure the term on left hand side being positive. 
Then, the OP can be given as
\begin{align}\label{q2 AF}
\nonumber
        q_2^{\mathrm{AF}}
        &= 1-\mathbb{P}( W_2 \geq 0, X\geq C_{x}, Z \geq C_{z})=1 -\int\limits_{C_0}^{\infty} \int\limits_{C_x}^{\infty}\big(1-\\  
        & \Phi_{\mathbf{J},2}(C_z) \big)d_{\mathrm{BF}}^{\alpha}e^{-d_{\mathrm{BF}}^\alpha{\mathit{x}}}   \ dx\  d_{\mathrm{BU}_2}^{\alpha}e^{-d_{\mathrm{BU}_2}^\alpha{\mathit{w}_2}}\ d{w_2},
\end{align}

Based on \eqref{q1 AF} and \eqref{q2 AF}, OP of user $2$ when using AF can be mathematically given as
    \begin{align}\label{AF OP}
    q^{\mathrm{AF}}=\left\{
    \begin{aligned}
    &q_1^{\mathrm{AF}},\ \mathrm{if\ \beta_{\mathrm{B}}<\frac{1}{1+\gamma_{\mathrm{U}_2}}} \\
    & q_2^{\mathrm{AF}},\ \mathrm{if\ \frac{1}{1+\gamma_{\mathrm{U}_2}} \leq \beta_{\mathrm{B}}<\frac{2}{2+\gamma_{\mathrm{U}_2}}}\\
    & 1,\ \mathrm{otherwise}
    \end{aligned}
    \right..
    \end{align}

\subsection{OP with DF Relaying}
As we have assumed, if use DF for relaying, the FAR can always decode $x_2$ from the received signal from BS, thus we have the following conclusion as 
\begin{equation}\label{X DF}
    X \geq \frac{\gamma_{\mathrm{U_2}}}{\xi_{\mathrm{B}}\rho_{\mathrm{B}}} = C_{w_2},
\end{equation}
from which we can also know $\xi_{\mathrm{B}}$ is positive, i.e., $\beta_{\mathrm{B}}<\frac{1}{1+\gamma_{\mathrm{U}_2}}$.

OP of the user $2$ can be expressed as in \eqref{q DF}, where $I_1 = \rho_{\mathrm{B}}[\beta_{\mathrm{F}}\big(1-\beta_{\mathrm{B}}\big)+\xi_{\mathrm{F}}\beta_{\mathrm{B}}] > 0 $ and $\hat{\gamma}_{\mathrm{U}_2} = \gamma_{\mathrm{U}_2} - W_2\xi_{\mathrm{B}}\rho_{\mathrm{B}}$.

\begin{figure*}
\begin{align}\label{q DF}
    \nonumber
&q^{\mathrm{DF}} 
    = 
    1 - \Big[  \mathbb{P}(\gamma_{\mathrm{FU}_1}^{\mathrm{1}}<\gamma_{\mathrm{U}_1}|\gamma_{\mathrm{FU}_2}^{\mathrm{1}} \geq \gamma_{\mathrm{U}_2}, \Gamma_{\mathrm{U}_2}^{\mathrm{DF-2}}\geq\gamma_{\mathrm{U}_2})+ \mathbb{P}(\gamma_{\mathrm{FU}_1}^{\mathrm{1}} \geq \gamma_{\mathrm{U}_1}|\gamma_{\mathrm{FU}_2}^{\mathrm{1}} \geq \gamma_{\mathrm{U}_2}, \Gamma_{\mathrm{U}_2}^{\mathrm{DF-F}}\geq\gamma_{\mathrm{U}_2}
    ) \Big]\\
    &=1- [ \underbrace{ \mathbb{P}(X<D_{w_1}|X\geq C_{w_2})\mathbb{P}(Z\geq \frac{\gamma_{\mathrm{U}_2} - W_2\xi_{\mathrm{B}}\rho_{\mathrm{B}}}{\rho_{\mathrm{F}}(W_2\rho_{\mathrm{B}}\beta_{\mathrm{B}}+1)})}_{\triangleq p_1^{\mathrm{DF}}} + 
    \underbrace{\mathbb{P}(X \geq D_{w_1}|X\geq C_{w_2})\mathbb{P}(Z\rho_{\mathrm{F}}(W_2\rho_{\mathrm{B}}I_1+ \xi_{\mathrm{F}}) \geq \hat{\gamma}_{\mathrm{U}_2}}_{\triangleq p_2^{\mathrm{DF}}}],
\end{align}
    
\hrulefill
\end{figure*}

Define $D_{w_1} = \frac{\gamma_{\mathrm{U}_1}}{\rho_{\mathrm{B}}\beta_{\mathrm{B}}}$ and $D_z = \frac{\gamma_{\mathrm{U}_2} - w_2\xi_{\mathrm{B}}\rho_{\mathrm{B}}}{\rho_{\mathrm{F}}(w_2\rho_{\mathrm{B}}\beta_{\mathrm{B}}+1)}$, we can find that when $D_{w_1} \leq C_{w_2}$, $p_{1}^{\mathrm{DF}} = 0$. When $D_{w_1} > C_{w_2}$, $p_{1}^{\mathrm{DF}}$ is 
\begin{align}
\nonumber
    &p_{1}^{\mathrm{DF}} =\big(1-e^{d_{\mathrm{BF}}^\alpha (C_{w_2} - D_{w_1})}\big)[e^{-d_{\mathrm{BU}_2}^\alpha C_{w_2}} \\
    &+ \int\limits_{0}^{C_{w_2}}\big(1- \Phi_{\mathbf{J},2}(C_z)\big)d_{\mathrm{BU}_2}^\alpha e^{-d_{\mathrm{BU}_2}^\alpha w_2}\ dw_2].
\end{align}

To derive the expression of $p_2^{\mathrm{DF}}$, we define $\xi_{\mathrm{F}} = 1 - \beta_{\mathrm{F}} -\beta_{\mathrm{F}}\gamma_{\mathrm{U}_2}$ and $D_{z'} = \frac{\gamma_{\mathrm{U}_2}-w_2\xi_{\mathrm{B}}\rho_{\mathrm{B}}}{\rho_{\mathrm{F}}\{w_2\rho_{\mathrm{B}}[\beta_{\mathrm{F}}(1-\beta_{\mathrm{B}})+\xi_{\mathrm{F}}\beta_{\mathrm{B}}]+\xi_{\mathrm{F}}\}}$. Then, when $D_{w_1} > C_{w_2}$, i.e., $\beta_{\mathrm{B}} < \frac{\gamma_{\mathrm{U}_1}}{\gamma_{\mathrm{U}_1}+\gamma_{\mathrm{U}_2}+\gamma_{\mathrm{U}_1}\gamma_{\mathrm{U}_2}}$, we can obtain that
\begin{align}
\nonumber
    &p_2^{\mathrm{DF}} = e^{d_{\mathrm{BF}}^\alpha (C_{w_2} - D_{w_1})} [e^{-d_{\mathrm{BU}_2}^\alpha C_{w_2}}+ \\
    & \int\limits_{\mathrm{max}\{0,\frac{-\xi_{\mathrm{F}}}{I_1}\}}^{C_{w_2}}\big(1-\Phi_{\mathbf{J},2}(D_{z'})\big)d^{\alpha}_{\mathrm{BU_2}}e^{-d^{\alpha}_{\mathrm{BU_2}}w_2}dw_2].
\end{align}
When $D_{w_1} \leq C_{w_2}$, i.e., $\beta_{\mathrm{B}} \geq \frac{\gamma_{\mathrm{U}_1}}{\gamma_{\mathrm{U}_1}+\gamma_{\mathrm{U}_2}+\gamma_{\mathrm{U}_1}\gamma_{\mathrm{U}_2}}$, we have
\begin{align}
\nonumber
   & p_2^{\mathrm{DF}} 
    = e^{-d_{\mathrm{BU}_2}^\alpha C_{w_2}} +\\
    & \int\limits_{\mathrm{max}\{0,\frac{-\xi_{\mathrm{F}}}{I_1}\}}^{C_{w_2}}\big(1-\Phi_{\mathbf{J},2}(D_{z'})\big)d^{\alpha}_{\mathrm{BU_2}}e^{-d^{\alpha}_{\mathrm{BU_2}}w_2}dw_2.
\end{align}

As a result, we can obtain $q^{\mathrm{DF}}$ as given in \eqref{DF OP}.
\begin{figure*}
    \begin{align}\label{DF OP}
    q^{\mathrm{DF}}=\left\{
    \begin{aligned}
    &1- e^{-d_{\mathrm{BU}_2}^\alpha C_{w_2}} - \big(1-e^{d_{\mathrm{BF}}^\alpha (C_{w_2} - D_{w_1})}\big) \int\limits_{0}^{C_{w_2}}\big(1- \Phi_{\mathbf{J},2}(C_{z})d_{\mathrm{BU}_2}^\alpha e^{-d_{\mathrm{BU}_2}^\alpha w_2}\ dw_2\\
       &-\int\limits_{\mathrm{max}\{0,\frac{-\xi_{\mathrm{F}}}{I_1}\}}^{C_{w_2}}e^{d_{\mathrm{BF}}^\alpha (C_{w_2} - D_{w_1})}\big(1- \Phi_{\mathbf{J},2}(D_{z'})\big)d^{\alpha}_{\mathrm{BU_2}}e^{-d^{\alpha}_{\mathrm{BU_2}}w_2}dw_2,\ \mathrm{if\ \beta_{\mathrm{B}}<\frac{\gamma_{\mathrm{U}_1}}{\gamma_{\mathrm{U}_1}+\gamma_{\mathrm{U}_2}+\gamma_{\mathrm{U}_1}\gamma_{\mathrm{U}_2}}}\\
    & 1 - e^{-d_{\mathrm{BU}_2}^\alpha C_{w_2}} - \int\limits_{\mathrm{max}\{0,\frac{-\xi_{\mathrm{F}}}{I_1}\}}^{C_{w_2}}\big(1-\Phi_{\mathbf{J},2}(D_{z'})\big)d^{\alpha}_{\mathrm{BU_2}}e^{-d^{\alpha}_{\mathrm{BU_2}}w_2}dw_2,\  \mathrm{if}\ \frac{\gamma_{\mathrm{U}_1}}{\gamma_{\mathrm{U}_1}+\gamma_{\mathrm{U}_2}+\gamma_{\mathrm{U}_1}\gamma_{\mathrm{U}_2}} \leq \beta_{\mathrm{B}} < \frac{1}{1+\gamma_{\mathrm{U_2}}}\ \\
    &  1, \ \mathrm{otherwise}
    \end{aligned}
    \right.
    \end{align}
    
\hrulefill
\end{figure*}

\subsection{Relaying Scheme Selection Principle}
From \eqref{AF OP} and \eqref{DF OP}, we can observe that the position of the FAR and power allocation coefficients all influence the OP of the transmission, and different relaying schemes also obtain different OP. To guarantee and improve the OP QoS of weak user $2$, the FAR always chooses the relaying scheme that maintains a lower transmission OP for user $2$. 

We define $\mu$ as the relaying scheme selection parameter as 
\begin{equation}\label{mu selection}
     \mu = \varepsilon(q^{\mathrm{AF}} - q^{\mathrm{DF}}),
\end{equation}
where $\varepsilon(\cdot)$ is the Heaviside function, specified as $\varepsilon(a)=1,\ \mathrm{when}\ a \geq 0$ and $\varepsilon(a)=0,\ \mathrm{when}\ a<0$. 

\section{Simulation Results}
Without loss of generality, we assume equal transmit powers at the BS and the FAR and equal OP QoS of ground users\cite{yue2017exploiting}. In particular we set transmit powers as $P_\mathrm{B}=P_{\mathrm{F}}$, i.e., $\rho_\mathrm{B}=\rho_{\mathrm{F}}=\rho$, and $\gamma_{\mathrm{U}_1} = \gamma_{\mathrm{U}_2}=\gamma$. The noise power is set as the $\sigma^2 = -130\ \mathrm{dBm}$. The FAS structure parameters are configured as $N_1=N_2=4$ and $L_1=L_2=1$. In our simulations, we first present the results on the OP of user $2$ under given different power allocation ratio coefficients by comparing our proposed scheme (labeled as `\textbf{Proposed}') with the scheme only deploying the ground link between BS and users (labeled as `\textbf{Without FAR}'). Then, we evaluate the value of $\mu$ under different FAR positions and outage thresholds.

\begin{figure*}[t]
    \centering
    \subfigure[]{
        \begin{minipage}{0.33\textwidth}
            \centering
            \includegraphics[width=1\textwidth]{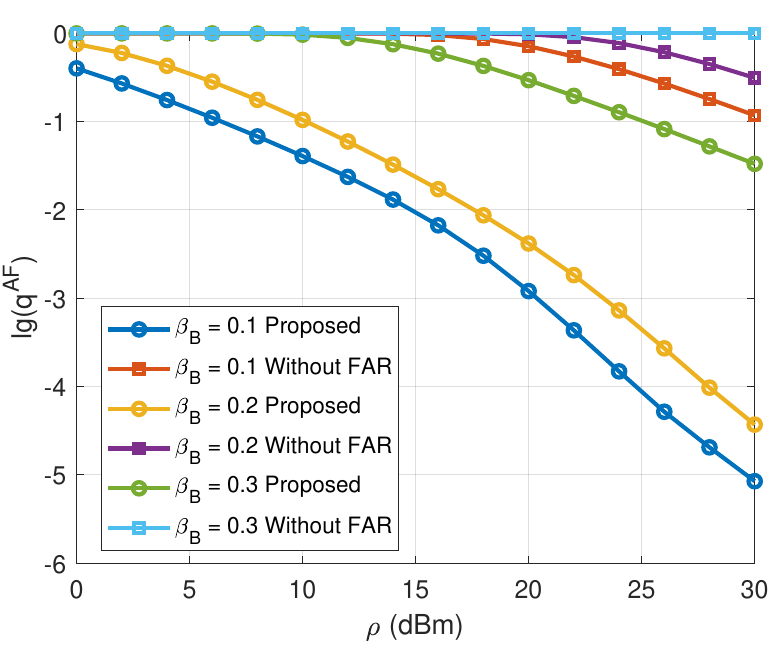}
            \label{Fig AF OP}
    \end{minipage}}
    \hspace{-5mm}
    \subfigure[]{
        \begin{minipage}{0.33\textwidth}
            \centering
            \includegraphics[width=1\textwidth]{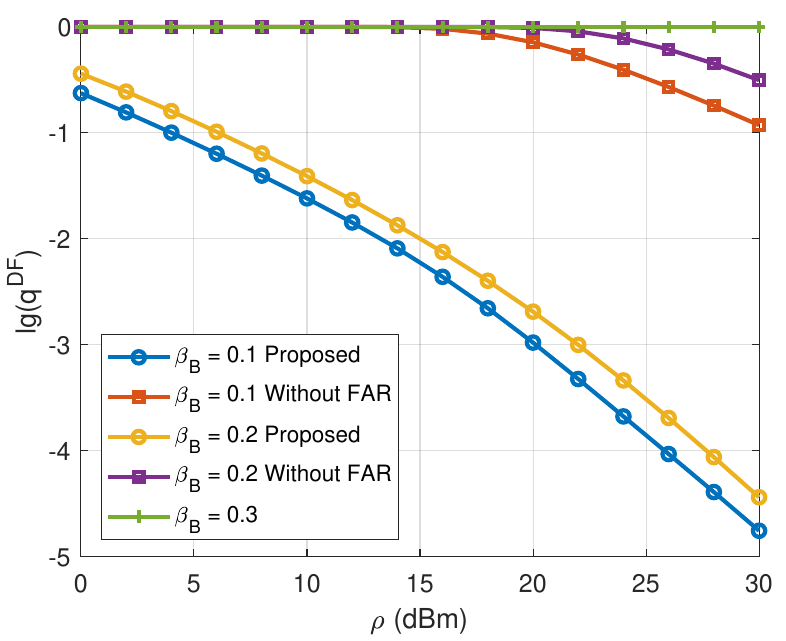}
            \label{Fig DF OP}	
    \end{minipage}}
        \hspace{-5mm}
        \subfigure[]{
            \begin{minipage}{0.33\textwidth}
                \centering
                \includegraphics[width=1\textwidth]{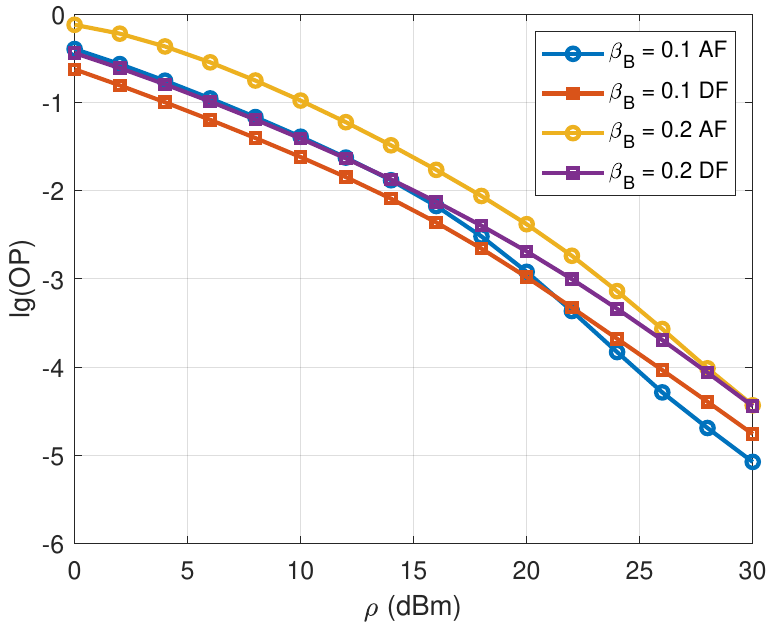}
                \label{Fig OP AFandDF}	
        \end{minipage}}
            \vspace{-0.5em}
    \caption{OP of user $2$ when the FAR uses (a) AF for relaying, (b) DF for relaying, and (c) different relaying schemes with different power allocation ratios. }
    \label{OP rho} 
\end{figure*}

Fig.~\ref{OP rho} shows the OP of user $2$ when the FAR uses AF or DF to relay with different power allocation ratio coefficients, where we set the outage threshold as $\gamma_{\mathrm{U}_2} = 3$. As illustrated in Fig.~\ref{Fig AF OP}, we evaluate the OP of user $2$ when the FAR uses AF. As the transmit powers increase, the OP values of all schemes decrease. When $\beta_{\mathrm{B}} = 0.1$ or $\beta_{\mathrm{B}} = 0.2$, the OP of the schemes transmitting without FAR first equals one and ranges within $10^{-1} \sim 10^{0}$ as SNR the system becomes sufficiently high. When the SNR is low, the OP of our proposed schemes falls in the order of magnitude similar to that of the comparison schemes mentioned earlier for high SNR. Meanwhile, the OP of the proposed schemes is less than $10^{-4}$, demonstrating a much better performance than the schemes without FAR. When $\beta_{\mathrm{B}} = 0.3$, which is close to the upper bound $\frac{2}{2+\gamma_{\mathrm{U}_2}}$, the scheme without FAR is always in outage as the OP equals $1$ and our proposed scheme can work with the OP falling in $10^{-2} \sim 10^{-1}$. Fig.~\ref{Fig DF OP} shows the OP of user $2$ when the FAR uses DF for relaying, with changing trends of the OP similar to those of using AF. Different from using AF, OP of user $2$ when using DF always equals 1 when $\beta_{\mathrm{B}} = 0.3$, since the power allocation ratio is greater than the bound $\frac{1}{1+\gamma_{\mathrm{U}_2}}$. To better illustrate the relationship between the OP values of AF and DF under given conditions, we represent Fig.~\ref{Fig OP AFandDF}. We can observe that when the SNR is low, the OP of the DF scheme with the same power allocation coefficient is smaller than that of the AF scheme, while at higher SNRs, the OP of the DF scheme becomes larger than that of the AF scheme. Furthermore, changes in the power allocation coefficient also affect these conclusions. For example, when $\beta_{\mathrm{B}}=0.2$, the OP of the DF scheme is larger than the OP of the AF scheme with $\beta_{\mathrm{B}}=0.1$ at low SNRs, rather than smaller.
\begin{figure*}[t]
    \centering
    \subfigure[]{
        \begin{minipage}{0.33\textwidth}
            \centering
            \includegraphics[width=1\textwidth]{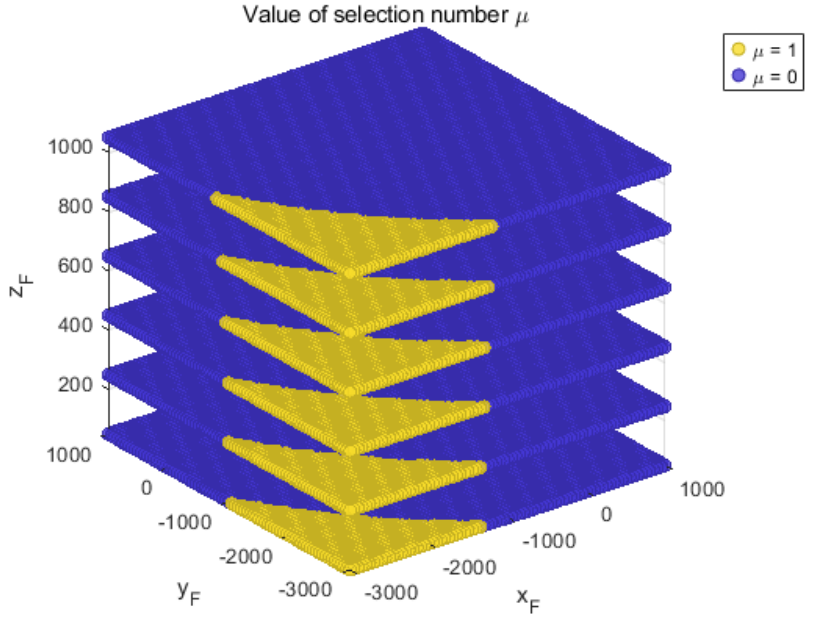}
            \label{muk 2.6}
    \end{minipage}}
    \hspace{-5mm}
    \subfigure[]{
        \begin{minipage}{0.33\textwidth}
            \centering
            \includegraphics[width=1\textwidth]{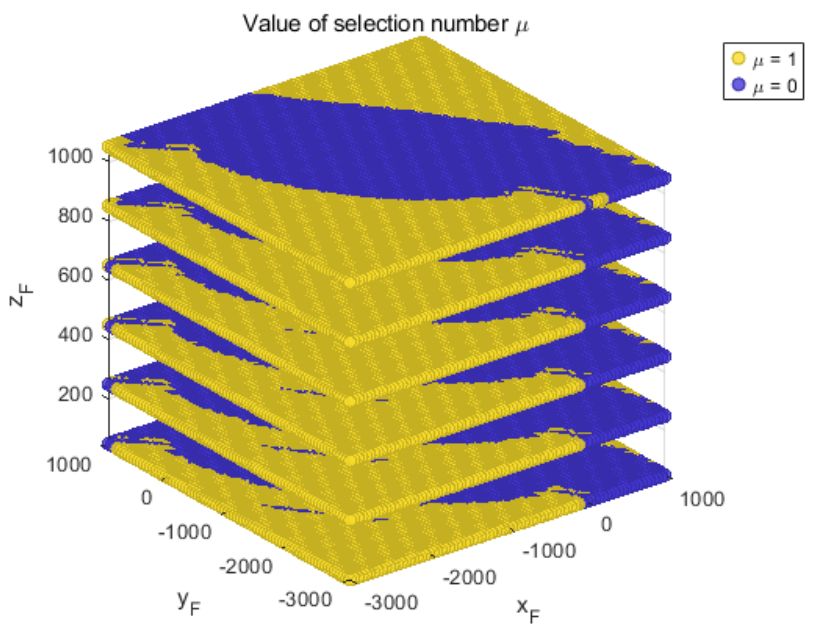}
            \label{muk 2.8}	
    \end{minipage}}
        \hspace{-5mm}
        \subfigure[]{
            \begin{minipage}{0.33\textwidth}
                \centering
                \includegraphics[width=1\textwidth]{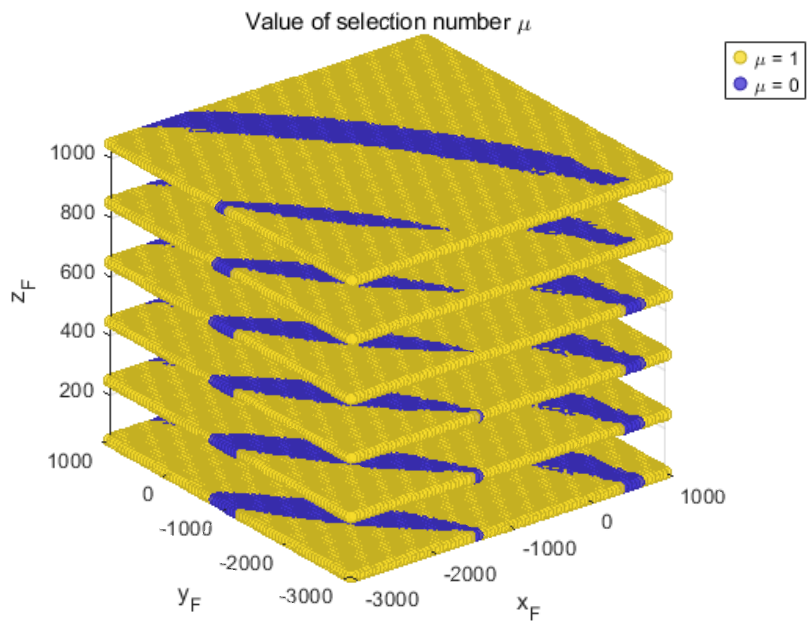}
                \label{muk}	
        \end{minipage}}
          \vspace{-0.5em}
    \caption{Value of $\mu$ with the FAR deployed at different feasible positions under the outage threshold given as (a) $\gamma_{\mathrm{U}_2} = 2.6$, (b) $\gamma_{\mathrm{U}_2} = 2.8$, and (c) $\gamma_{\mathrm{U}_2} = 3$. }
    \label{mukValue} 
\end{figure*}

Fig.~\ref{mukValue} depicts the values of $\mu$ as FAR is at different positions when three different outage thresholds are given, i.e., $\gamma_{\mathrm{U}_2} = 2.6$, $\gamma_{\mathrm{U}_2} = 2.8$ and $\gamma_{\mathrm{U}_2} = 3$. In general, we set the power allocation ratio as $\beta_\mathrm{B} = \beta_\mathrm{F} = 0.1$, and the value of $\mu$ varies depending on the outage threshold and the position of FAR, where $\mu $ is calculated based on \eqref{mu selection}. Furthermore, as the FAR height increases, the area where $\mu$ equals 1 gradually decreases when $\gamma_{\mathrm{U}_2} = 2.6$ and increases when $\gamma_{\mathrm{U}_2} = 2.8$ and when $\gamma_{\mathrm{U}_2} = 3$. Fig.~\ref{muk 2.6} shows the scenario when $\gamma_{\mathrm{U}_2} = 2.6$. It can be observed that, under these circumstances, when the FAR is deployed in most areas, we should use AF for transmission to achieve a lower OP for user $2$. When the FAR is located far from the BS and close to user $2$, we should use DF for transmission because the channel gain $h_{\mathrm{FU}_2}$ increases, resulting in a larger probability of $p_2^{\mathrm{DF}}$ than corresponding part of AF. This conclusion also holds true for the other two threshold values. The situation changes as the outage threshold increases. As shown in Fig.~\ref{muk 2.8}, when the FAR is close to user 1, $\mu$ changes from value of $0$ to $1$, indicating that the forwarding strategy at the same location may change as outage threshold increases or decreases. As $\gamma_{\mathrm{U}_2}$ further increases to $3$, we can observe in Fig.~\ref{muk} that, at the same height, the area near the central where AF is used for forwarding further decreases. However, compared to when $\gamma_{\mathrm{U}_2} = 2.8$, a new area near user 2 appears where AF forwarding should be used. The change in $\mu$ reflects the highly nonlinear nature, making the analysis of OP particularly important.

\section{Conclusion}
In this paper, we have analyzed the performance of an FAR-assisted downlink NOMA communication system and derived the OP expressions for the user maintaining weaker channel conditions under both AF and DF relaying schemes. Simulation results demonstrate that deploying the FAR significantly reduces the OP compared to transmission without the FAR, especially when the channel conditions are poor. With given FAR position and outage threshold, the choice between AF and DF depends on parameters such as transmit power ratio and the SNR. In particular, at lower SNRs with given simulation conditions, the DF scheme typically yields a lower OP than the AF scheme with the same power allocation, a trend that reverses at higher SNRs. In addition, numerical simulations also show that the optimal relaying selection strategy is highly dependent on the position of FAR and the outage threshold, confirming that the relaying strategy can be dynamically adjusted accustomed to the communication conditions.

\bibliographystyle{IEEEtran}
\bibliography{MMM}

\end{document}